\documentclass[prb,twocolumn,showpacs,floatfix,amsmath,amssymb,superscriptaddress]{revtex4-1}

\usepackage{amsfonts}
\usepackage{stmaryrd}
\usepackage{bbm}
\usepackage{mathrsfs}
\usepackage{tipa}
\usepackage{amssymb}
\usepackage{txfonts}
\usepackage{graphicx}
\usepackage{dcolumn}
\usepackage{epstopdf}
\usepackage[colorlinks,linkcolor=blue,urlcolor=blue,citecolor=blue]{hyperref}
\usepackage{multirow}
\usepackage{subfigure}
\usepackage{url}
\usepackage{setspace}

\begin{document}

\title{Optical spectroscopy and ultrafast pump-probe study of structural phase transition in 1T'-TaTe$_2$ }
\author{T. C. Hu}
\affiliation{International Center for Quantum Materials, School of Physics, Peking University, Beijing 100871, China}

\author{Q. Wu}
\affiliation{International Center for Quantum Materials, School of Physics, Peking University, Beijing 100871, China}

\author{Z. X. Wang}
\affiliation{International Center for Quantum Materials, School of Physics, Peking University, Beijing 100871, China}

\author{L. Y. Shi}
\affiliation{International Center for Quantum Materials, School of Physics, Peking University, Beijing 100871, China}

\author{Q. M. Liu}
\affiliation{International Center for Quantum Materials, School of Physics, Peking University, Beijing 100871, China}

\author{L. Yue}
\affiliation{International Center for Quantum Materials, School of Physics, Peking University, Beijing 100871, China}

\author{S. J. Zhang}
\affiliation{International Center for Quantum Materials, School of Physics, Peking University, Beijing 100871, China}

\author{R. S. Li}
\affiliation{International Center for Quantum Materials, School of Physics, Peking University, Beijing 100871, China}

\author{X. Y. Zhou}
\affiliation{International Center for Quantum Materials, School of Physics, Peking University, Beijing 100871, China}

\author{S. X. Xu}
\affiliation{International Center for Quantum Materials, School of Physics, Peking University, Beijing 100871, China}

\author{D. Wu}
\affiliation{International Center for Quantum Materials, School of Physics, Peking University, Beijing 100871, China}
\affiliation{Songshan Lake Materials Laboratory, Dongguan, Guangdong 523808, China}

\author{T. Dong}
\affiliation{International Center for Quantum Materials, School of Physics, Peking University, Beijing 100871, China}

\author{N. L. Wang}
\email{nlwang@pku.edu.cn}
\affiliation{International Center for Quantum Materials, School of Physics, Peking University, Beijing 100871, China}
\affiliation{Beijing Academy of Quantum Information Sciences, Beijing 100913, China}

\begin{abstract}

 1T'-TaTe$_2$ exhibits an intriguing first-order structural phase transition at around 170 K. Understanding the electronic structural properties is a crucial way to comprehend the origin of the structural phase transition. We performed a combined optical and ultrafast pump-probe study on the compound across the transition temperature. The phase transition leads to abrupt changes of both optical spectra and ultrafast electronic relaxation dynamics. The measurements revealed a sudden reconstruction of band structure. We elaborate that the phase transition is of the first order and can not be attributed to the conventional density-wave type instability. Our work is illuminating for understanding the origin of the structural phase transition.
\end{abstract}

\pacs{Valid PACS appear here}

\maketitle

\section{INTRODUCTION}
Transition-metal dichalcogenides (TMDS) have attracted tremendous interests for their intriguing ground states, physical properties and potential applications \cite{doi:10.1080/00018736900101307,Ali2014,Yu2015}. Those low-dimensional compounds of these systems often lead to charge density wave (CDW) instability \cite{doi:10.1080/00018737500101391,Morosan2006a,PhysRevLett.92.086401,PhysRevB.76.045103,PhysRevLett.102.166402}.  Yet, many of those CDW-bearing materials are superconducting \cite{VanMaaren1967}. The interplay between the two very different cooperative electronic phenomena is one of the pending issues in condensed-matter physics. The origin of  CDW is most commonly traced to fermi surface nesting. However, the driving force 
for higher dimensional CDW formation in compounds are rather complex so that many mechanisms were constantly proposed \cite{PhysRevLett.35.120,PhysRevLett.86.4382,Kiss2007,PhysRevB.77.165135,PhysRevLett.101.226406,Rossnagel2011}. The compounds such as well-known Ta/NbX$_2$ (X=S, Se, Te) with 1T and 2H sandwich structure provide fundamental insight into the behavior of correlated electron systems. Among the TaX$_2$ (X=S, Se, Te) polytypes, the telluride TaTe$_2$ exhibits two 
intriguing commensurate CDW states but has not been well studied by spectrocopy yet. Compare with octahedral polymorphs 1T-TaS$_2$ and 1T-TaSe$_2$, 1T’-TaTe$_2$ presents stronger electron-phonon interaction and larger lattice distortion. The ground state of TaTe$_2$ is metallic rather than Mott insulating. The trimerization of Ta atoms in 1T structure lead to double zigzag chains and an overall monoclinic distortion of the crystal lattice. So TaTe$_2$ can be regarded as distorted TaTe$_6$ octahedra 1T’ structure with 3×1 “ribbon-chain” clusters superstructure at room temperature. Undergoing a structural phase transition at around 170 K, TaTe$_2$  transforms into 3×3 “butterfly-like” clusters superstructure\cite{Landuyt1974,SORGEL2006987}. The transition leads to anomaly in specific heat, magnetic susceptibility and electrical resistivity (distinct kink structure with hysteresis)\cite{SORGEL2006987,Liu_2015,Chen_2017}. No different polytypes with 1T and 2H structures have been found for TaTe$_2$ as so far. The low temperature (LT) phase could be fully suppressed and a new superconducting phase emerges at T$_c^{onset}$ $\sim$ 1.7-2.5 K  when Tellurium (Te) is partially replaced by Selenium (Se) \cite{PhysRevB.94.045131,LuoE1174}. Additionally, applying external pressure could achieved the bulk superconductivity of TaTe$_2$ between 4-6 K in the pressure range of 21-50 GPa\cite{guo2017separation}. As for isostructural CDW compound 1T’-NbTe$_2$ with 3×1 superstructure at 300 K\cite{Brown:a04997}, such a LT phase has not been observed but exhibits superconductivity at T$_c$  $\sim$ 0.5 K\cite{Zhang2019}. Another analogue layered compound IrTe$_2$ with a 5d heavy transition metal element and strong Te-Te interlayer interaction shows a structural phase transition and transport properties anomaly as well \cite{PhysRevLett.108.116402}. Whereas the absence of gap opening in IrTe$_2$ demonstrated by both optical study and angle resolved photoemission spectroscopy (ARPES), chemical bonding states and crystal field splitting theory were considered as origins of the transition\cite{RN7,Ootsuki2013}.

It is essential to know the origin across different CDW states in two-dimensional CDW systems because of their close connection to superconductors. Up to now, few theoretical works suggested that the origin of the CDW states in TaTe$_2$ were driven by fermi surface nesting\cite{PhysRevB.66.195101,PhysRevB.97.045133}. On the other hand, limited experiment works mainly focused on the measurements of the transport and structure properties\cite{Liu_2015,Chen_2017,PhysRevB.98.195423,PhysRevB.98.224104,PhysRevB.102.024111,PhysRevLett.125.165302,PhysRevB.103.064103}. Transport measurements suggested possible occurrence of fermi surface reconstruction just below T$_s$\cite{Liu_2015,Chen_2017}. However, recently Kar et al pointed that the fermi surface topology hardly changes across T$_s$ by their ARPES study on TaTe$_2$\cite{Kar2021}. The available experiment results were still controversial. Other spectroscopy technique detections are still lacking. It is well-known that optical and ultrafast spectroscopy are sensitive probes for bulk electronic properties of solids. Therefore, it is significant to uncover the nature of the structural phase transition by above techniques.

In this work, we performed combined temperature-dependent optical spectroscopy and ultrafast pump-probe measurements on single-crystal 1T’-TaTe$_2$  sample . The optical reflectivity spectra suddenly change just below T$_s$ over a broad energy range, yielding evidence for reconstruction of fermi surface and an explicit reduction of conducting carriers in the LT phase. However, no characteristic feature of energy gap opening related to CDW formation is observed. Ultrafast pump-probe measurement reveals dramatic change in the electronic relaxation dynamics near T$_s$. With the temperature increasing across T$_s$, the amplitude of transient reflectivity is enhanced enormously and the decay time becomes much faster. Furthermore, a number of coherent phonons in LT phase were extracted by reflectivity oscillations in time domain. Nevertheless, no amplitude mode of CDW collective excitation could be identified. In combination with above experiment results we elaborate that the structural phase transition is not of a density wave type and can not be simply attributed to conventional fermi surface nesting driven instability. Our work is illuminating for understanding the transition of the different CDW states.

\section{RESULTS AND DISCUSSION}

\begin{figure}[htbp]
	\centering
	\includegraphics[width=9cm]{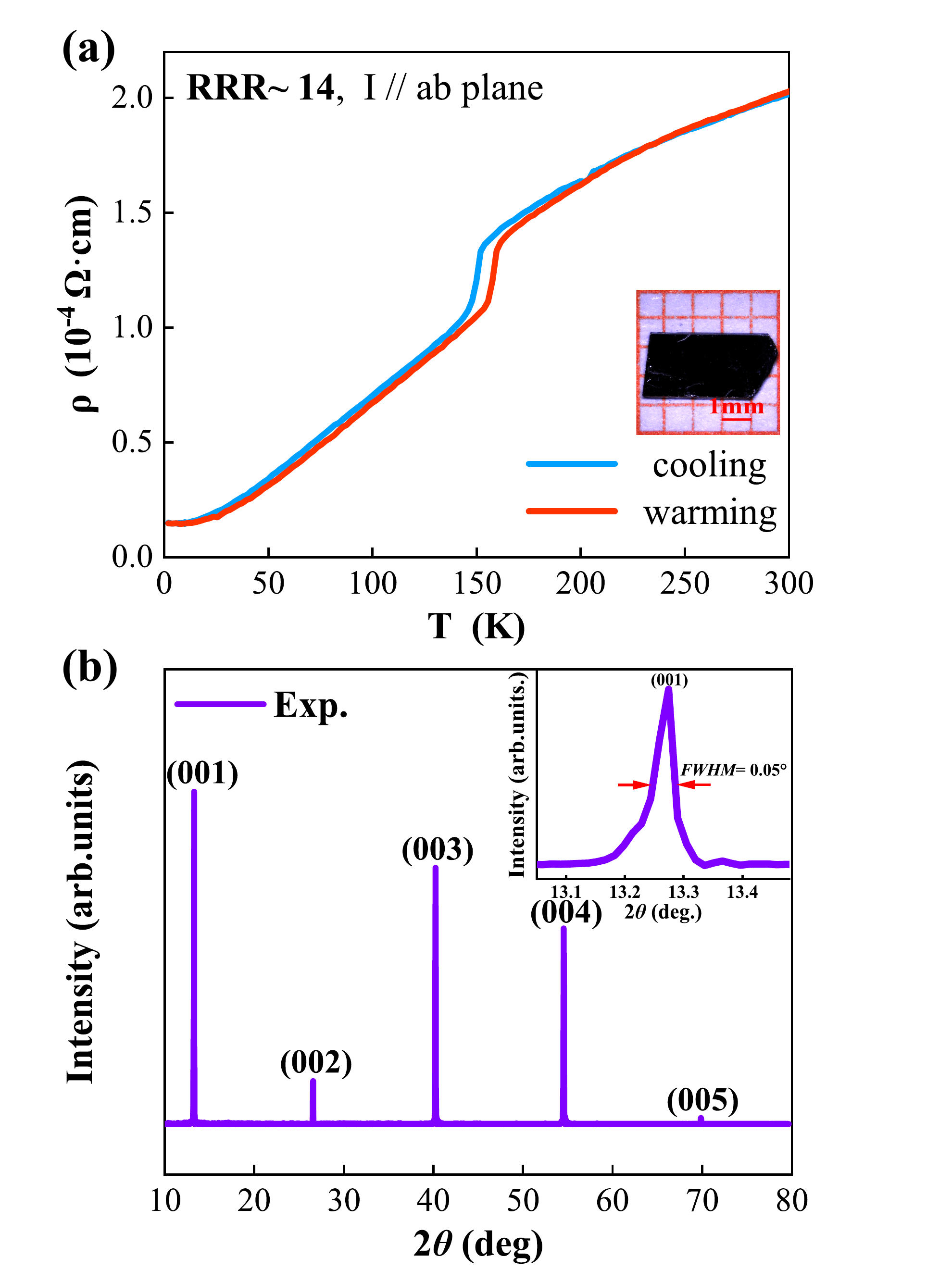}\\
	\caption{\textbf{Sample characterizations for TaTe$_2$.} (a) Temperature-dependent ab-plane resistivity measurement, a structural phase transition is evident near 160 K. Inset: Single crystal picture of TaTe$_2$. (b) Room temperature powder XRD patterns of single-crystal and indexing. Inset: The full width at half maximum (FWHM) of (001) brag peak is about 0.05°, indicating good quality of single crystal.
}\label{Fig:1}
\end{figure}

The single crystals of TaTe$_2$ were synthesized by chemical vapor transport method with iodine as transport agent. High-purity Ta powders (99.99\%) and Te pellets (99.999\%) in a stoichiometric ratio (1:2) with additional iodine (99.8\%, 5 mg/cm$^3$) were loaded into a quartz tube. All processes were operated in a sealed glove box. Then the quartz tube was sealed under high vacuum and heated at 560 $^{\circ}$C - 480 $^{\circ}$C in a two-zone furnace\cite{PhysRevB.98.224104}. After 10 days growth, shiny platelike single crystals (typical size of 4×2.5×0.02 mm$^3$, inset of Fig.\ref{Fig:1} (a)) were obtained. The temperature-dependent resistivity was measured by a standard four-probe method with the cooling/warming rate of 2K/min in the ab plane. The measurement was performed in a Quantum Design physical property measurement system (PPMS). Resistivity measurements, shown in Figure \ref{Fig:1}(a), indicating metallic behavior with a small resistivity value of about 2.1×10$^{-4}$ $\Omega\cdot$cm at 300 K. Similar to many two-dimensional CDW systems, TaTe$_2$ remains metallic state at LT phase and the residual resistance ratio $\left( R R R=\rho_{300 \mathrm{~K}} / \rho_{2 \mathrm{~K}}\right)$ is about 14. The resistivity curve shows a significant kink hysteresis at around 160 K which suggests the phase transition is of the first order. X-ray diffraction (XRD) experiment results at room temperature are shown in Fig.\ref{Fig:1} (b). The strong diffraction peaks of TaTe$_2$ crystals can be indexed as (00L) of monoclinic phase of TaTe$_2$ (ICDD-PDF21-1201). It can be seen that the full width at half maximum (FWHM) of the (001) Bragg peak is only about 0.05° in the inset of Fig.\ref{Fig:1} (b), indicating the high quality of TaTe$_2$ single crystal. Detailed chemical compositions investigated by energy dispersive spectroscopy (EDS) measurements show that the chemical atomic ratios are close to the standard stoichiometric ratio (Ta:Te $\sim$ 1:1.96). All these sample fundamental characterizations are in good agreement with reports in literature\cite{SORGEL2006987,PhysRevB.98.224104,PhysRevB.103.064103}.

\begin{figure}[htbp]
	\centering
	\includegraphics[width=9cm]{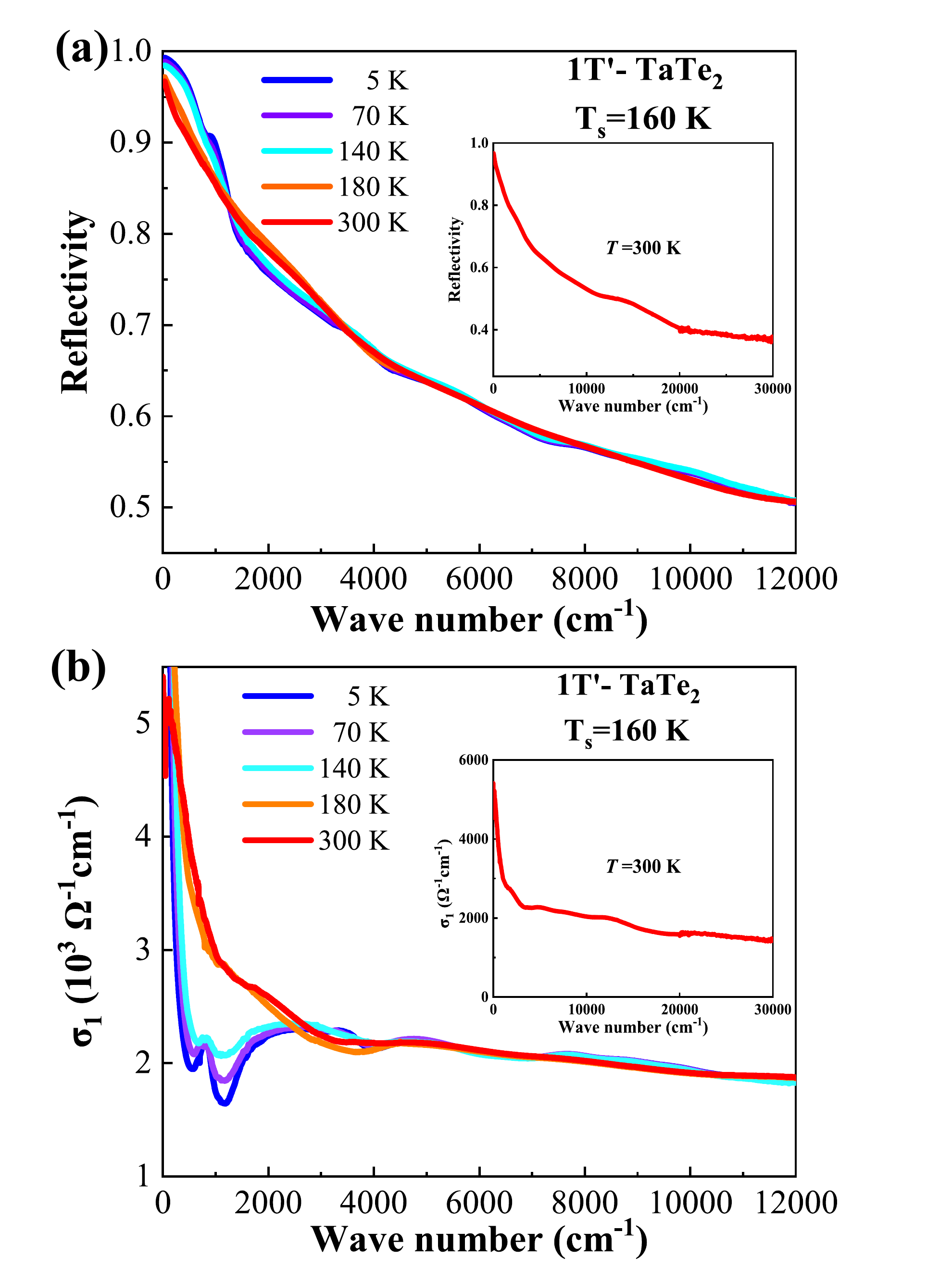}\\
	\caption{\textbf{Temperature-dependent optical spectroscopy of  TaTe$_2$.} (a) temperature-dependent optical reflectivity measurements below 12000 cm$^{-1}$, Inset: Large energy scale range of 100-30000 cm$^{-1}$ at 300 K. (b) temperature-dependent optical conductivity below 12000 cm$^{-1}$,Inset: Optical conductivity spectrum with large energy scale range of 100-30000 cm$^{-1}$ at 300 K.
}\label{Fig:2}
\end{figure}

The as grown ab-plane (001) optical reflectance measurements were performed on the Fourier transform infrared spectrometer Bruker 80V in the frequency range from 100 to 30000 cm$^{-1}$. An in-situ gold and aluminum evaporation technique was used to get the reflectance $R(\omega)$. The main panel of Fig.\ref{Fig:2} (a) shows the reflectivity up to 12000 cm$^{-1}$ at five selected temperatures. The inset displays the experimental reflectance spectrum up to 30000 cm$^{-1}$ at 300 K. A good metallic response is observed in the $R(\omega)$ spectrum: At low-frequency, $R(\omega)$ has high values and approaches unit at zero frequency limit. Below 4000 cm$^{-1}$, the frequency-dependent reflectivity decreases almost linearly in the high temperature (HT) phase. This behavior is similar to the high-temperature cuprate superconductors. $R(\omega)$ shows minor change as temperature decreases from 300 K to 180 K. However, when temperature decreases just below the structural phase transition temperature at 160 K, optical reflectivity $R(\omega)$ shows an abrupt change. With further decreasing temperature, the spectral change becomes very small. The very low frequency $R(\omega)$ increases slightly and forms a small plateau in LT phase, reflecting enhanced metallic dc conductivity and consisting with the above resistivity measurements. In the meantime, two weak suppressions in the mid-infrared region near 800 cm$^{-1}$ and 2300 cm$^{-1}$ could be illustrated.

The real part of optical conductivity $\sigma_1(\omega)$ was derived from $R(\omega)$ through Kramers-Kronig transformation in Fig.\ref{Fig:2} (b). The Hagen-Rubens relation was used for the low-energy extrapolation of $R(\omega)$. We have employed the x-ray atomic scattering functions in the high-energy side extrapolation\cite{PhysRevB.91.035123}. The main panel of Fig.\ref{Fig:2} (b) displays $\sigma_1(\omega)$ below 12000 cm$^{-1}$  at five selected temperatures, the inset displays $\sigma_1(\omega)$ up to 30000 cm$^{-1}$  at 300 K. The Drude-type conductivity was observed in all spectra at low frequency. In HT phase, the broad width of Drude peak indicates a large scattering rate of the itinerant carriers. Upon entering the LT phase, the weight of Drude-type conductivity was suddenly removed. The optical spectra kept almost unchanged at 10 K and 70 K. The overall spectral change reflects a significant band structure reconstruction associated with the structural phase transition. In other words, there are two totally different metallic states in HT and LT phase.

Actually, for some structural phase transitions that are relevant to CDW order formation have some expected optical spectroscopic features: the formation of an energy gap with related spectral change only at low energy. The so-called case-\uppercase\expandafter{\romannumeral1} factor of CDW condensate would cause a sharp continuous rise in the optical conductivity spectral just above the energy gap. With temperature further decreasing the dip feature in reflectivity becomes more and more prominent. These features were observed definitely in most of the typical conventional CDW systems driven by fermi surface nesting such as rare-earth tri-telluride (RTe$_3$)\cite{PhysRevB.90.085105}, LaAgSb$_2$\cite{PhysRevLett.118.107402}, Bi$_2$Rh$_3$Se$_2$\cite{PhysRevB.101.205112} and CuTe\cite{li2021optical}. On the contrary, for a purely structural phase transition that is irrelevant to CDW order formation, the band structure of HT and LT phase is entirely different. So optical spectra would change suddenly over a broad range energy scale as observed in BaNi$_2$As$_2$\cite{PhysRevB.80.094506}, IrTe$_2$\cite{RN7} and RuP\cite{PhysRevB.91.125101}. For TaTe$_2$, the spectral change occurs in a wide frequency range and no continually evident reflectivity suppression is observed. The overall spectral change is similar to the latter characteristics which are referred to the purely structural phase transition.

\begin{figure}[htbp]
	\centering
	\includegraphics[width=9cm]{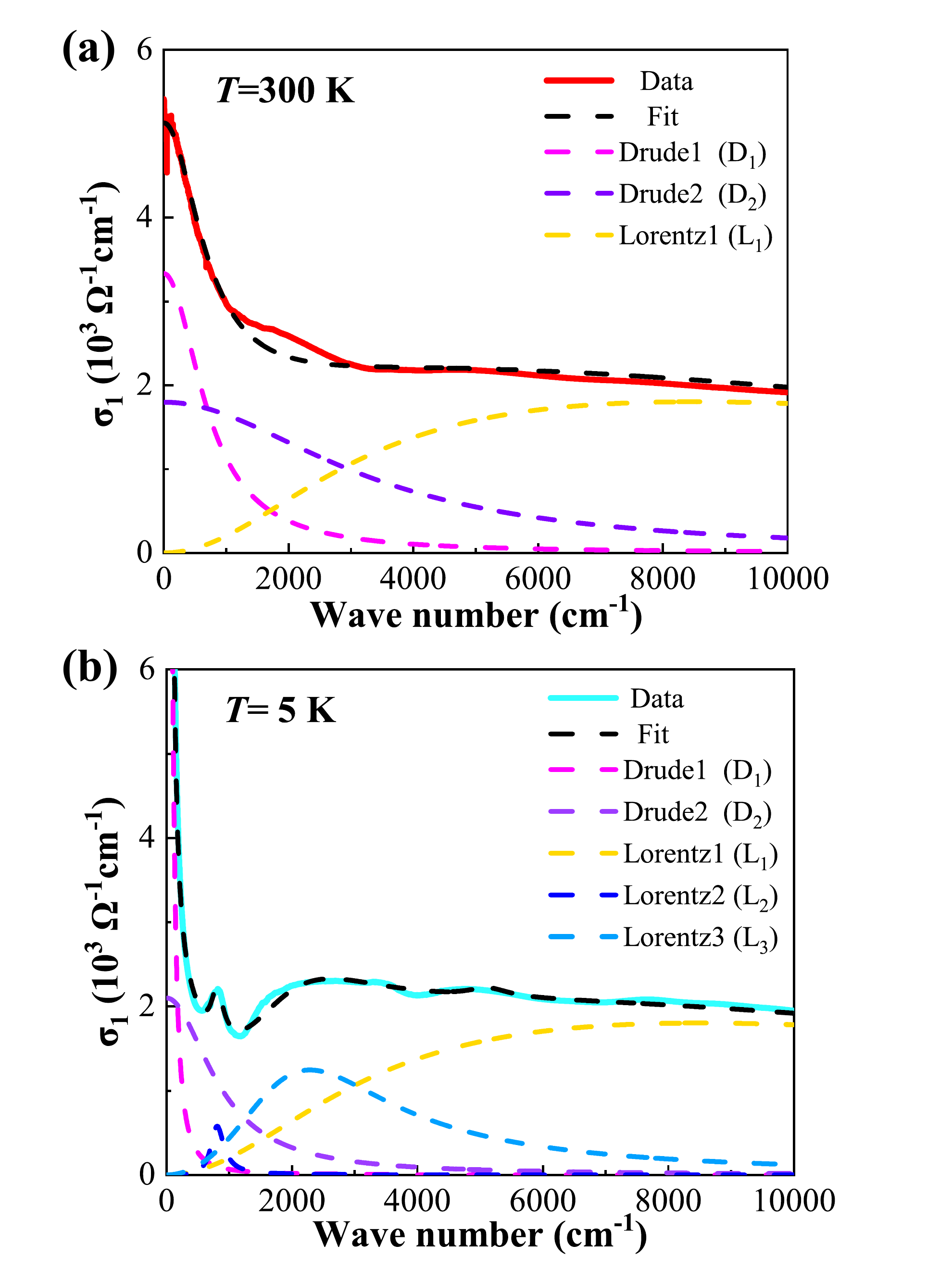}\\
	\caption{\textbf{Drude-Lorentz fitting of optical conductivity  for TaTe$_2$.} (a) $T$=300 K. (b) $T$=5 K.
}\label{Fig:3}
\end{figure}

In order to isolate the different components of the electronic excitations and make quantitative analysis, we employ the Drude-Lorentz model to decompose the optical conductivity. The Drude component represents the contribution from conduction electrons while the Lorentz components are used to describe the excitations across energy gaps and interband transitions. The general formula for the Drude-Lorentz model is \begin{equation}
\sigma_{1}(\omega)=\sum_i \frac{\omega_{p i}^{2}}{4 \pi} \frac{\Gamma_{D i}}{\omega^{2}+\Gamma_{D i}^{2}}+\sum_j \frac{S_{j}^{2}}{4 \pi} \frac{\Gamma_{j} \omega^{2}}{\left(\omega_{j}^{2}-\omega^{2}\right)^{2}+\omega^{2} \Gamma_{j}^{2}}
\end{equation}

where $\omega_{p i}$ and $\Gamma_{D i}$ are the plasma frequency and the relaxation rate of each conduction band while $\omega_{j}$, $\Gamma_{j}$, and $S_{j}$ represents resonance frequency, the damping, and the mode strength of each Lorentz oscillator, respectively. In general, a single Drude term is used to extract the itinerant electron contribution but we found that two Drude components could reproduce the low frequency conductivity much better. This could be attributed to the multi-band characteristic of the TaTe$_{2}$ naturally\cite{PhysRevB.94.045131}. Similar two Drude component analysis has been applied to the optical data of 1T'-MoTe$_{2}$ and T$_{d}$ -WTe$_{2}$\cite{PhysRevB.99.195203}, pointing toward a generic behavior for those multi-band systems. The general formula for the Drude-Lorentz model is reproduced by two Drude components and one Lorentz oscillator at 300 K. Below the transition temperature, two additional Lorentz oscillators (Lorentz 2 and Lorentz 3) are required to fit the curve. Fig.\ref{Fig:3} (a) and (b) shows the spectra and fitting curves at two typical temperatures 300 K and 5 K, respectively. The second Drude term is much broader and has much larger spectral weight than the first one. In LT phase at 5 K, both Drude terms shrink substantially and two more Lorentz (L2 and L3) oscillations centered at around 800 cm$^{-1}$ and 2300 cm$^{-1}$ could be resolved.

The general plasma frequency is obtained as $\omega_{p}=\left(\omega_{p 1}^{2}+\omega_{p 2}^{2}\right)^{1 / 2}$ for overall perspective, the calculated $\omega_{p}$ decreases from 22800 cm$^{-1}$ to 15200 cm$^{-1}$ across phase transition. The overall plasma frequency indicates that a majority of itinerant charge carriers are suddenly reduced across the transition. In the meantime, as evidenced by the narrowing of the Drude peak, both of the scattering rates drops violently, from 720 cm$^{-1}$ to 32 cm$^{-1}$ for Drude 1 term, from 3310 cm$^{-1}$ to 854 cm$^{-1}$ for Drude 2 term. The sharply decrease of scattering rate makes even lower DC resistivity in LT phase despite the partial free carriers removed. The measurement reveals a sudden reconstruction of band structure.

\begin{figure*}[htbp]
	\centering
	\includegraphics[width=15cm]{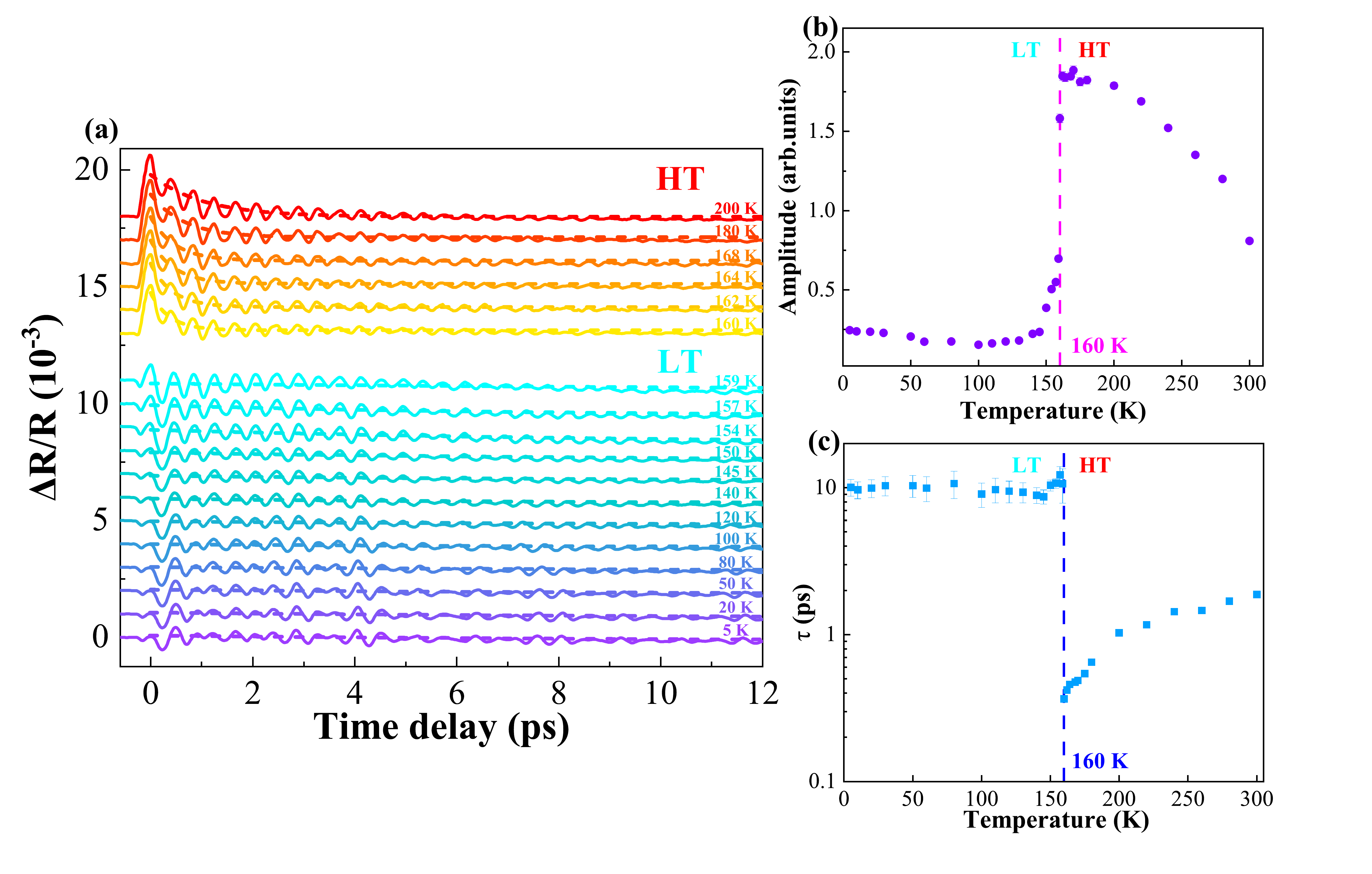}\\
	\caption{\textbf{Temperature-dependent ultrafast relaxation dynamics for TaTe$_2$.} (a) $\Delta R/R$ in the temperature range of 5-200 K at some selected temperature points.  The dash lines are the single-exponential fitting curves: $\Delta R/R(t) = A {\cdot}\rm{e}^{-t/\tau_d}+{C}.$ The amplitude A (purple cycles) in (b) and decay time in (c) (blue squares) of the transient reflectivity extracted from fits to the single-exponential above and below structural phase transition T$_s$ $\approx$ 160 K. Error bars represent the standard deviation of the fit. 
}\label{Fig:4}
\end{figure*}

Our ultrafast pump-probe experiment could further support that the structural phase transition is not of density wave type. The ultrafast pump-probe experiment with standard reflection geometry configuration was performed in a Ti:sapphire amplified laser (Spitfire Ace) of 800-nm pulses with 35-fs duration at 1-KHz repetition frequency. To minimum sample heating effect as possible, the fluence of pump beam is set to around 90 $\mu $ J/cm$^{2}$ while the probe beam is thirty times weaker than the pump beam. The spot size was determined to be 110 and 90 $\mu$m in diameter by a 100 $\mu$m diameter pinhole for the pump and probe spots, respectively. The pump and probe pulses were set to cross-polarized and an extra grid-polarizer was just mounted before the silicon detector (Thorlabs, DET36A2) in order to reduce the noise from stray light. And the probe pulses were splitted for balanced detection in order to improve signal noise ratio.

Figure \ref{Fig:4}(a) presents the photo-induced transient reflectivity as a function of time delay on a freshly-cleaved TaTe$_2$ at a number of selective temperatures. All data were collected when the sample was warmed up from low temperature. A sudden change was found precisely at the phase transition temperature (T$_s$ $\approx$ 160 K). This sudden electron relaxation dynamics change was also observed just after structural phase transition in YbInCu$_4$ and CsV$_3$Sb$_5$\cite{PhysRevB.95.165104,PhysRevB.104.165110}. Very prominently, with temperature increasing across T$_s$, the amplitude of the photo-induced $\Delta R /R$ is enhanced dramatically. In LT phase, the pump pulse induces a small rise followed by a relative slow decay process in transient reflectivity while there is a fast decay dynamics at HT phase. The measurement reveals that the compound exhibits two quietly different electronic states just above and below the T$_s$.

In order to quantitatively analyze the relaxation of the photo-excited quasiparticles, we use a single-exponential function to fit the decay process: $\Delta R / R=A \exp \left(-t / \tau_{d}\right)+C$, Where $A$ in the formula represents the amplitude of the photo-induced reflectivity change, $\tau_{d}$ stands for the relaxation time and $C$ is constant for long lived thermal diffusion process. The best fitting curves are shown as dashed lines in Fig.\ref{Fig:4} (a). The temperature dependence of the extracted fitting parameters  $A$  and  $\tau_{d}$ of the single-exponential function are shown in Fig.\ref{Fig:4} (b) and (c). It is evident that the compound shows totally different behaviors above and below T$_s$. In the HT phase, the decay time of relaxation process gradually increases from $\sim$ 0.4 ps at 160 K to $\sim$ 1.5 ps at 300 K, but becomes almost temperature-independent in the LT phase with much slower process at around 10 ps. As for the amplitude of reflectivity, the value varies a little until the temperature rises close to T$_s$. With temperature changes across T$_s$, the amplitude suddenly increases from $\sim$ 0.4 at 150 K to $\sim$ 1.6 at 160 K and the whole increasing process is a typical step shape like.

\begin{figure*}[htbp]
 \centering
 \centering\includegraphics[width=18cm]{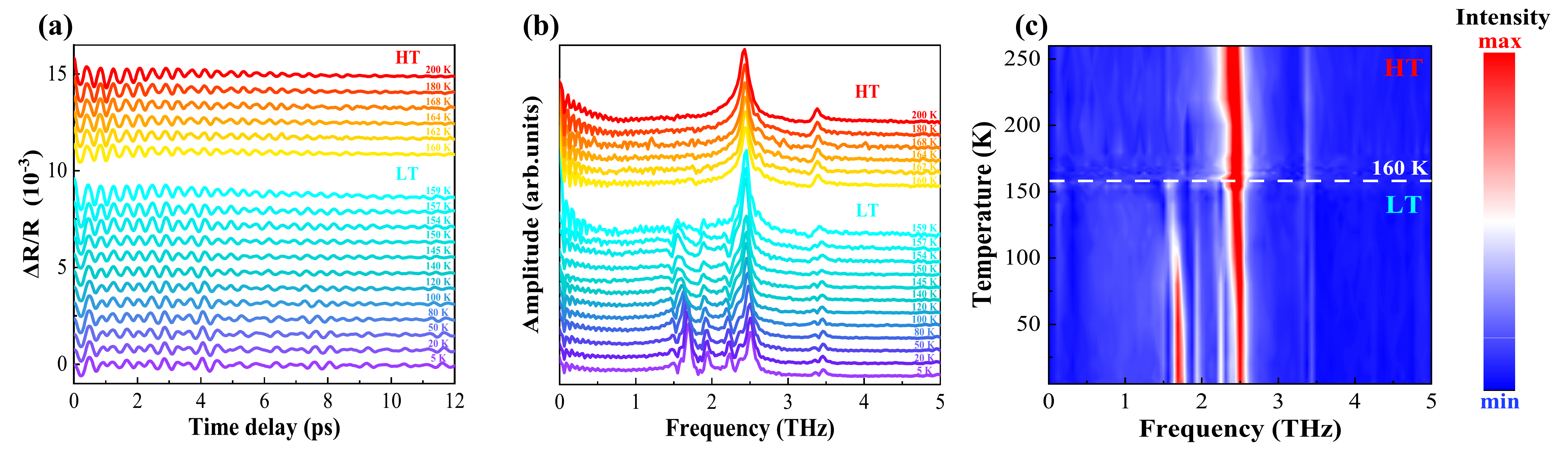}\\
 \caption{
\textbf{Temperature-dependent coherent phonon spectroscopy for TaTe$_2$ in time/frequency domain.} (a) Coherent phonon oscillations results in time-domain after the decay background is subtracted. (b) The fast fourier transformation (FFT) results of the data in (a). (c) Temperature-dependent intensity map extracted from (b).} \label{Fig:5}
\end{figure*}

Ultrafast pump-probe spectroscopy could also provide coherent phonon oscillations’ information after subtracting the electron dynamics background. We obtained the coherent phonon oscillations in Fig.\ref{Fig:5} (a) and fast fourier transform (FFT) results are displayed in Fig.\ref{Fig:5} (b). A detailed two-dimensional intensity map as a function of temperature and frequency is presented in Fig.\ref{Fig:5} (c). The coherent phonon frequencies are in agreement with the previous Raman experiment results on TaTe$_2$\cite{PhysRevB.103.064103}. At HT phase, there are two distinct peaks at 2.4 THz and 3.3 THz in the frequency domain. At LT phase, more phonon modes abruptly appear due to the symmetry lowering caused by Ta atoms further clustering, yielding evidence for the new structural modulation. Remarkably, phonon at 3.3 THz even splits into one A$_g$ mode phonon peak and one B$_g$ mode phonon peak which is in good agreement with the density functional theory (DFT) calculations for the LT phonon spectra\cite{PhysRevB.98.224104}. The distinct phonon mode splitting near 3.3 THz further indicates that the phase transition is a first-order phase transition. On the other hand, conventional CDW condensate also has collective excitations referred to as amplitude mode and phase mode. Usually, the energy level of amplitude mode is in the range of terahertz frequency which could be identified by the ultrafast pump-probe experiment\cite{PhysRevLett.118.107402,PhysRevB.101.205112,li2021optical,PhysRevLett.83.800,PhysRevB.66.041101,PhysRevLett.101.246402}.  The amplitude mode exhibits more pronounced oscillations with temperature decreasing and usually appears as the strongest oscillations. The mode frequency softens and behaves like an order parameter when temperature gets close to the T$_{CDW}$\cite{PhysRevLett.118.107402}. Nevertheless, no such CDW amplitude mode could be identified from the above phonon spectra at LT phase. Similarly, we noticed that amplitude mode was also not observed in recently reported Kagome metal CsV$_3$Sb$_5$ coherent phonon spectra in LT CDW phase \cite{Ratcliff2021,PhysRevB.104.165110}. It could also suggest that the structural phase transition is not of conventional density wave type from another aspect.

The above pump-probe experiment results reveal a rather specific situation: There are two entirely different behaviors above and below structural phase transition temperature. Both the transient reflectivity amplitude and decay time change dramatically. A number of coherent phonon modes suddenly appear in LT phase without clear trace of softening at the structural phase transition temperature. No observed phonon modes in the LT phase phonon spectra could be identified as CDW amplitude mode. All those features are well consistent with the first-order phase transition and entirely different from a usual conventional second-order CDW condensate behaviors.

Based on our experiment results, we could infer that the origin of structural phase transition at T$_s$ $\approx$ 160 K could not be attributed to the fermi surface nesting like many typical two-dimensional CDW systems, i.e., RTe$_2$ and RTe$_3$ (R=rare earth metal) \cite{PhysRevLett.98.166403,PhysRevB.90.085105}. Apparently, further theoretical calculations and analyses are still needed to elucidate the driving force of the structural phase transition in 1T’-TaTe$_2$ precisely.

\section{SUMMARY}
In summary, we have performed a combined optical spectroscopy and ultrafast pump-probe study on single-crystal 1T’-TaTe$_2$ in an effort to understand the electronic structure in the LT phase. The optical study revealed that the electronic band structure reconstruction over a broad energy scale. And time-resolved pump-probe measurements provide further evidence for the phase transition is of first order that irrelevant to the conventional CDW order formation. Both studies indicated that the structural phase transition was not driven by the conventional CDW instability. Our research on 1T’-TaTe$_2$ would motivate further investigations on the origin or driving force of the phase transition.

\begin{center}
\small{\textbf{ACKNOWLEDGMENTS}}
\end{center}

This work was supported by the National Key Research and Development Program of China (No. 2017YFA0302904), the National Natural Science Foundation of China (No. 11888101).

\bibliography{TaTe2}

\end{document}